\documentclass{epl}
\usepackage[english]{babel}
\usepackage{graphicx}
\usepackage{dcolumn}
\usepackage{bm}

\makeatletter
\def\vereq#1#2{%
 \lower3\p@\vbox{%
  \baselineskip1.5\p@
  \lineskip1.5\p@
  \ialign{$\m@th#1\hfill##\hfil$\crcr#2\crcr\sim\crcr}%
 }%
}%
\makeatother
\newcommand{\lesssim}{\mathrel{\mathpalette\vereq<}}
\newcommand{\gtrsim}{\mathrel{\mathpalette\vereq>}}

\newcommand{\mathi}{{\rm i}}
\newcommand{\diff}{{\rm d}}

\title{\centerline{Tomonaga--Luttinger parameters for doped Mott insulators}}

\shortauthor{Satoshi Ejima et al.}
\shorttitle{Tomonaga--Luttinger parameters} 

\author{Satoshi Ejima\inst{1} \and Florian Gebhard\inst{1} \and 
Satoshi Nishimoto\inst{2}}
\institute{\inst{1} Fachbereich Physik, Philipps-Universit\"at Marburg,
D-35032 Marburg, Germany\\
\inst{2} Institut f\"ur Theoretische Physik, Universit\"at G\"ottingen,
D-37077 G\"ottingen, Germany}


\begin{document}

\maketitle

\begin{abstract}%
The Tomonaga--Luttinger parameter $K_{\rho}$ determines
the critical behavior in quasi one-dimensional correlated
electron systems, e.g., the exponent $\alpha$ for the density of states
near the Fermi energy.
We use the numerical
density-matrix renormalization group method to calculate~$K_{\rho}$
from the slope of the density-density correlation function
in momentum space at zero wave vector.
We check the accuracy of our new approach against exact results for the
Hubbard and XXZ Heisenberg models.
We determine $K_{\rho}$ in the phase diagram of the extended Hubbard model
at quarter filling, $n_{\rm c}=1/2$, and confirm 
the bosonization results $K_{\rho}=n_{\rm c}^2=1/4$ on the critical line
and $K_{\rho}^{\rm CDW}=n_{\rm c}^2/2=1/8$ at infinitesimal doping 
of the charge-density-wave (CDW) insulator for all interaction strengths. 
The doped CDW insulator exhibits exponents~$\alpha>1$ 
only for small doping and strong correlations.
\end{abstract}

Quasi one-dimensional correlated 
electronic systems offer the unique opportunity
of a detailed comparison between theory and experiment.
Some models for correlated lattice electrons can be solved exactly,
e.g., the XXZ~Heisenberg chain and the Hubbard model, and many of 
the physical properties of Hubbard-type models
are known from analytical and numerical studies.
Moreover, their low-energy properties turn out to be universal,
i.e., they belong to the generic class of Tomonaga--Luttinger liquids (TLL).
It has been a challenge to detect the signatures of TLL in experiments
on metallic single-walled carbon nanotubes~\cite{nano2004},
anisotropic transition-metal oxides like PrBa$_2$Cu$_4$O$_8$~\cite{Tak2000},
Bechgaard salts like (TMTSF)$_2$ClO$_4$~\cite{Bour19841993},
and TTF-TCNQ~\cite{Sing2003}.

The low-energy properties of TLL are characterized
by few quantities, most notably the TL~parameter~$K_{\rho}$
which determines the long-range behavior of the density-density
correlation function
and the exponent $\alpha$ for the density of states
$D(\omega)$ near the Fermi energy~\cite{Voit95},
\begin{equation}
D(\omega\to 0)\sim |\omega|^{\alpha} \quad , \quad 
\alpha = (K_{\rho}+K_{\rho}^{-1}-2)/4\; .
\end{equation}
A value $\alpha\gtrsim 1$ was estimated for TTF-TCNQ~\cite{Sing2003} 
and in early works on 
(TMTSF)ClO$_4$~\cite{Bour19841993}
which would imply 
$K_{\rho}\lesssim 3-2\sqrt{2}\approx 0.17$.
More recent measurements 
report~$K_{\rho}=0.23$ for (TMTSF)ClO$_4$~\cite{Vescoli1998} 
and $K_{\rho}=0.24$ for PrBa$_2$Cu$_4$O$_8$~\cite{Tak2000}.
In the Hubbard model, $K_{\rho}^{\rm H}\geq 1/2$ so that
$\alpha^{\rm H} \leq 1/8$. 
Therefore, if electronic correlations are indeed responsible
for the algebraic behavior of the density of states near the Fermi energy,
$\alpha>1/8$ is only possible when 
the long-range parts of the Coulomb interaction are taken seriously.
Unfortunately, the calculation of $K_{\rho}$ 
for correlated-electron models is very difficult. 

In this work we employ the density-matrix renormalization group (DMRG) method
for a reliable numerical calculation of the TL~parameter.
In order to demonstrate our approach
we consider $N$~interacting spin-1/2 electrons 
on a chain with an even number~$L$~sites. 
The electron density is
$n_{\uparrow}=n_{\downarrow}=n/2=N/(2L)$. 
In the absence of a Peierls dimerization, 
the extended Hubbard model provides an appropriate description
for correlated electron systems in one dimension,
\begin{equation}
\hat{H} = -t \sum_{l,\sigma} \left(
\hat{c}^+_{l,\sigma} \hat{c}_{l+1,\sigma} + 
\hat{c}^+_{l+1,\sigma} \hat{c}_{l,\sigma} \right)
+ U \sum_l \hat{n}_{l,\uparrow}\hat{n}_{l,\downarrow}
+ V \sum_{l}
\left(\hat{n}_{l} - n \right) \left(\hat{n}_{l+1} - n \right) \; , 
\label{eqn:hamiltonian}
\end{equation}
where $\hat{c}^+_{l,\sigma}$ ($\hat{c}_{l,\sigma}$) is the creation 
(annihilation) operator of an electron with spin 
$\sigma$ on site~$l$,  
$\hat{n}_{l,\sigma}=\hat{c}^+_{l,\sigma}\hat{c}_{l,\sigma}$
is the number operator, and
$\hat n_l=\hat n_{l,\uparrow}+\hat n_{l,\downarrow}$.
Moreover, $t$~is the electron transfer integral 
between neighboring sites, 
$U$~is the strength of the Hubbard interaction, and the nearest-neighbor
interaction~$V$ models the long-range part of the electrons' Coulomb repulsion.

The density-density correlation function 
is defined by the ground-state expectation value
\begin{equation}
C^{\rm NN}(r) 
= \frac{1}{L} \sum_{l=1}^{L} 
\langle \hat{n}_{l+r} \hat{n}_{l} \rangle
- \langle \hat{n}_{l+r} \rangle 
\langle \hat{n}_{l} \rangle \; ,
\label{CNNr}
\end{equation}
where $\hat{n}_l=\sum_{\sigma}\hat{n}_{l,\sigma}$
counts the electrons on site~$l$. 
We have $C^{\rm NN}(r)=C^{\rm NN}(-r)$ due to
inversion symmetry, and periodic boundary conditions apply.

Using conformal field theory it can be shown~\cite{Fra90,Sch90} that
the asymptotic behavior for $1\ll r\ll L$ is given by
\begin{equation}
C^{\rm NN}(r) \sim
-\frac{K_{\rho}}{(\pi r)^2}
+\frac{A\cos(2k_{\rm F}r)}{r^{1+K_{\rho}}}\ln^{-3/2}(r)
+\cdots\; ,
\label{eqn:den1}			 
\end{equation}
where $k_{\rm F}=n\pi/2$ is the Fermi wave number, and $A$~is a constant.
For spinless Fermions, the first term should be multiplied by~$1/2$.
Field theory further predicts~\cite{GiaSchulz}
$K_{\rho}(V_{\rm c}(U);n_{\rm c})=n_c^2$ 
for an interaction-driven metal-insulator
transition. For example, 
we have $K_{\rho}(V_{\rm c}(U);1/2)=1/4$
for the CDW transition in the quarter-filled extended Hubbard model. 
In contrast, for
the density-driven metal-insulator transition
the value from field theory is
$K_{\rho}(V>V_{\rm c}(U),n_{\rm c}^-)=n_c^2/2$, e.g.,
for the extended
Hubbard model field theory predicts 
$K_{\rho}(V>V_{\rm c}(U),1/2^-)=1/8$ in the
infinitesimally doped CDW phase at quarter filling.

As seen from the exactly solvable cases discussed below,
we expect that $K_{\rho}(U,V;n_c-\delta)$ can be expanded in a Taylor series
in the doping~$\delta=n_c-n\ll 1$,
\begin{equation}
K_{\rho}(U,V,n_c-\delta) = 
\frac{n_{\rm c}^2}{2} + \frac{\delta}{h(U,V)} + \cdots \; .
\label{Krhoexpansion}
\end{equation}
Moreover, since the critical line 
is a line of Kosterlitz--Thouless transitions, 
we expect that the prefactor diverges exponentially
in the vicinity of the critical line,
\begin{equation}
h(V\to V_{\rm c}(U)^+) \sim 
\exp\left( \frac{C}{(V-V_{\rm c}(U))^{\gamma}}\right) \; .
\label{TaylorKrhoCDW}
\end{equation}
Therefore, close to the critical line 
the value $K_{\rho}=n_{\rm c}^2/2$ of the CDW insulator 
cannot be observed for finite doping.

Field theory does not quantify the convergence radius of~(\ref{Krhoexpansion})
or the region in which~(\ref{TaylorKrhoCDW}) holds.
To answer this question, the TL~parameter
must be calculated for the extended Hubbard model
with accurate numerical methods.
We can extract $K_{\rho}$ via Fourier transformation,
\begin{equation}
\widetilde{C}^{\rm NN}(q) 
= \sum_{r=1}^{L} e^{-\mathi qr} C^{\rm NN}(r)  \; ,
\label{FT}			 
\end{equation}
with $0\leq q <2\pi$.
By construction, $\widetilde{C}^{\rm NN}(q=0)=0$. 
For the derivative at $q=0$ one finds in the thermodynamic 
limit~\cite{Dz95,Noa99}
\begin{equation}
K_{\rho} = \pi \lim_{q \to 0^+} \frac{\widetilde{C}^{\rm NN}(q)}{q}
\; .
 \label{eqn:den2}			 
\end{equation}
In numerical simulations we treat finite systems.
There, Eq.~(\ref{eqn:den2}) translates into
\begin{equation}
K_{\rho} = \lim_{L \to \infty} 
\frac{L}{2}\widetilde{C}^{\rm NN}\left(\frac{2\pi}{L}\right) \; .
\label{eqn:den3}			 
\end{equation}
Several groups have calculated the density-density correlation 
function in position space for Hubbard-type models.
After Fourier transformation they obtained $K_{\rho}$ 
from~(\ref{eqn:den3}), see, e.g., \cite{Dz95,Noa99,Dau98,Cla99}. 
The main problem of this approach lies in the accurate calculation
of~$C^{\rm NN}(r)$ from~(\ref{CNNr}) for large distances.
The accuracy of the correlation function 
becomes worse as the distance~$r$ increases
which severely limits the precision of 
the Fourier transform~$\widetilde{C}^{\rm NN}(q)$, especially
for small~$q$. 

In this work we 
calculate $\widetilde{C}^{\rm NN}(2\pi/L)$ directly in momentum space.
To this end we define 
\begin{equation}
N(q)  = \frac{1}{L}
\langle \Psi_0| \hat{n}(q) \hat{n}(-q)|\Psi_0 \rangle
\label{eqn:Nq2}			 
\end{equation}
for $q=2\pi m/L$ ($m\geq 1$),
where $\hat{n}(q)$ is given by
\begin{equation}
\hat{n}(q)=\hat{n}^+(-q)= \sum_{l,\sigma}
e^{-\mathi q(l-r_c)}\hat{c}^+_{l, \sigma}\hat{c}_{l,\sigma} \; .
 \label{eqn:nq}			 
\end{equation}
Here, $r_c=(L+1)/2$ denotes the central position of the chain. 
Note that $N(q)$ and $\widetilde{C}^{\rm NN}(q)$ are different.
It is only in the thermodynamic limit, when boundary effects are absent,
that they become identical. Therefore, 
\begin{equation}
K_{\rho}(L) = \frac{L}{2} N\left(\frac{2\pi}{L}\right)  \quad , \quad
K_{\rho} = \lim_{L \to \infty} K_{\rho}(L) \; . \label{neweqn:Nq2}
\end{equation}
The important idea is to target not only the ground state in 
the DMRG procedure but also
the state $|\Psi_q\rangle = \hat{n}(-q)|\Psi_0 \rangle$.
In this way, a precise DMRG calculation of~$N(q)$ 
and of~$K_{\rho}$ from~(\ref{neweqn:Nq2}) 
becomes possible.

We illustrate the accuracy of our method for the 
Hubbard model [$V=0$ in~(\ref{eqn:hamiltonian})],
for which $K_{\rho}(U;n)$ is known from the Bethe Ansatz solution~\cite{Sch90}.
We investigate systems with $L\leq 128$ 
and open boundary conditions. 
The number of density matrix states kept is $m=1500$, so that the 
maximum truncation error is $3 \times 10^{-6}$. 

\begin{figure}[htbp]
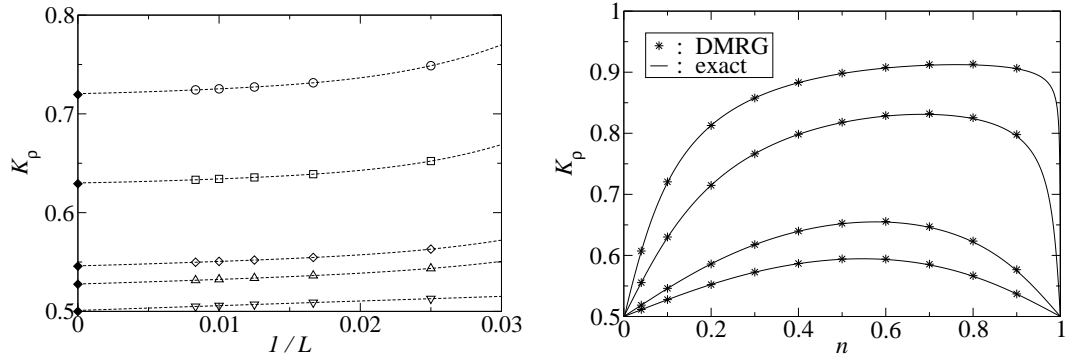

  \begin{center}
    \resizebox{6.8cm}{!}{\includegraphics{figs/extrap.eps}} \quad 
    \resizebox{6.8cm}{!}{\includegraphics{figs/tUmodel.eps}}
  \end{center}
\caption{Left: $K_{\rho}(L)$ 
as a function of the inverse system size
in the one-dimensional Hubbard model at $n=0.1$ 
for $U/t$=1 (circles), $U/t=2$ (squares), $U/t=6$ (diamonds), 
$U/t=10$ (upward triangles), and $U/t=\infty$ (downward triangles). 
           Diamonds are exact values from the Bethe Ansatz in the
           thermodynamic limit. Lines are 4th-order 
polynomial fits.
\newline
Right:
TL~parameter $K_{\rho}$ in the one-dimensional 
Hubbard model as a function of the density for 
for $U/t=1, 2, 6, 10$ (from top to bottom).
The full lines are exact values from the Bethe Ansatz,
stars mark the results from DMRG.\label{fig1}}
\end{figure}

In Fig.~\ref{fig1}a, we show 
$K_{\rho}(L)$ as a function of the inverse 
system size for several values of~$U/t$. 
The band filling is fixed at $n=0.1$, which, apart from the limit~$n\to 1$,
is the most difficult parameter region in this model because 
$K_{\rho}$ changes significantly as a function of the interaction strength.
For all values~$U>0$, $K_{\rho}(L)$ is found to decrease monotonically 
as a function of inverse system size, so that we can
extrapolate $K_{\rho}$ to the thermodynamic limit systematically by
performing a least squares fit of $K_{\rho}$ to a polynomial 
in~$1/L$. 

In Fig.~\ref{fig1}b we compare our results for $K_{\rho}^{\rm DMRG}$
with those from Bethe Ansatz for various fillings and interaction 
strengths. The relative error 
$|K_{\rho}^{\rm DMRG}-K_{\rho}^{\rm exact}|/K_{\rho}^{\rm exact}$ is 
below $0.3\%$ for all DMRG data shown.
We reproduce the exact results with a much better accuracy
than exact diagonalization~\cite{Sch90},
the DMRG method~\cite{Noa99}  
based on the Fourier transformation formula~(\ref{eqn:den2}),
and the calculation of $K_{\rho}$ from
the compressibility and the charge velocity~\cite{Eji04}.

\begin{figure}[htbp]
\vspace*{12pt}
  \begin{center}
    \resizebox{8.0cm}{!}{\includegraphics{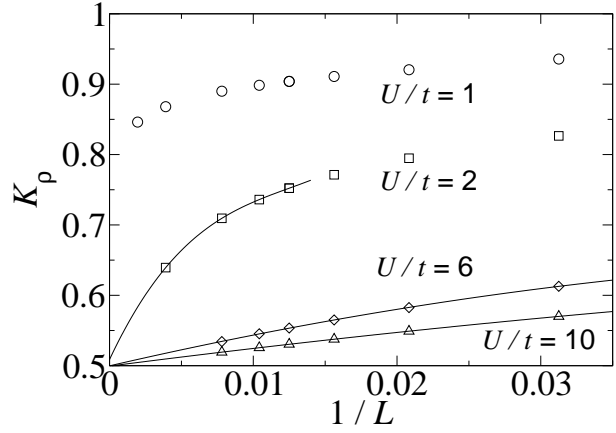}}
  \end{center}
  \caption{$K_{\rho}(L)$ as a function of the inverse system size
for the infinitesimally doped Mott--Hubbard insulator, $n=1-2/L$,
for various interaction strengths. It extrapolates to the exact value
$K_{\rho}=1/2$ for all~$U/t$. \label{newfig3}}
\end{figure}

For the Mott--Hubbard transition at $U_{\rm c}=0^+$
the Taylor expansion~(\ref{TaylorKrhoCDW}) applies~\cite{Kawa} ($n_{\rm c}=1$,
$f(U)=h(U,0)$)
\begin{equation}
f(U) = \frac{4}{\ln 2} \int_{1}^{\infty} \frac{\diff x}{\sqrt{x^2-1}}
\frac{1}{\sinh(2\pi tx/U)}\quad , \quad 
f(U\to 0)\sim \exp\left(-\frac{2\pi t}{U}\right) \; .
\label{TaylorKrho}
\end{equation}

Fig.~\ref{newfig3} shows our results for the
infinitesimally doped Mott--Hubbard insulator, $n_{\rm h}=2/L$,
for various values of the interaction strength.
The approach is valid because
we calculate $N(2\pi/L)$ so that the momentum~$q=2\pi/L$ transferred from
the ground state~$|\Psi_0\rangle$ to $|\Psi_q\rangle$ 
is of the same order of magnitude as the phase shifts induced by the
introduction of two holes.
For $U\gtrsim 2t$ our systems are larger than the correlation length
of the system, and
we reproduce $K_{\rho}(U>0,n\to 1)=1/2$ numerically.
Finite-size effects become prominent for $U\to 0$, 
and it is difficult to recover
the exact value from numerical calculations
in that limit.

\begin{figure}[htbp]
  \begin{center}
    \resizebox{6.8cm}{!}{\includegraphics{figs/tUVinset.eps}}
\quad 
\resizebox{6.282cm}{!}{\includegraphics{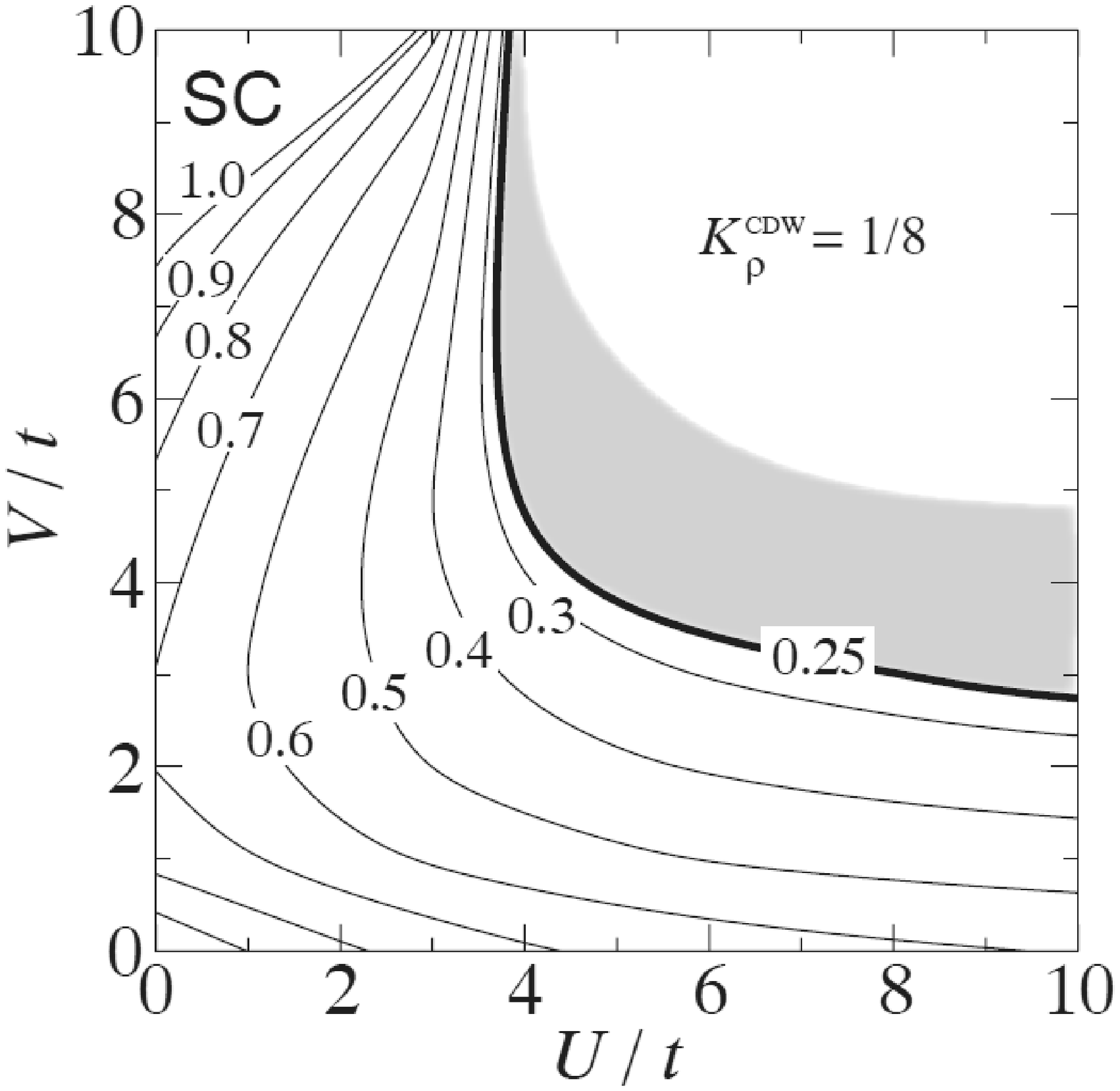}}
  \end{center}
  \caption{Left: TL~parameter $K_{\rho}$ for the extended
Hubbard model at $U/t=\infty$, $V\leq 2t$ from DMRG (stars) 
and exact Bethe Ansatz (full line).
Inset: $K_{\rho}(L)$ as a function of inverse system size
for periodic boundary conditions.
\newline
Right: Contour map for the TL~parameter $K_{\rho}$ in the $U$-$V$
 plane of the extended Hubbard model at quarter filling. 
The bold line represents the boundary of the metal-insulator 
transition. The infinitesimally doped charge-density-wave insulator
(CDW) has $K_{\rho}=1/8$. The shaded area indicates the region
with an exponentially small gap.\label{fig3}}
\end{figure}

As our second test we study the extended Hubbard model at quarter filling,
$n=1/2$, 
for $U = \infty$, which can be mapped onto
the exactly solvable Heisenberg XXZ~chain.
The parameter $K_{\rho}$ for the Hubbard model
at $U=\infty$ from the Bethe Ansatz is ($V\leq V_{\rm c}=2t$)
\begin{eqnarray}
 K_{\rho}=\frac{\pi}{4\arccos[-V/(2t)]} \; .
 \label{eqn:krho_spinless}			 
\end{eqnarray}
In Fig.~\ref{fig3}a, we show $K_{\rho}$ as a function of~$V$ for
$U=\infty$ together with the exact result. 
For this system we use periodic boundary conditions 
because $m=2000$ density-matrix eigenstates are enough 
to calculate $K_{\rho}(L)$ with high precision, and 
finite-size effects are much smaller for periodic than for open boundary 
conditions,
see the inset in Fig.~\ref{fig3}a.
Relative errors 
$|K_{\rho}^{\rm DMRG}-K_{\rho}^{\rm exact}|/K_{\rho}^{\rm exact}$
are below $0.5\%$, even for $V=1.95t$ where the system
is close to the CDW insulator.
Again, 
the predictions~(\ref{Krhoexpansion}), (\ref{TaylorKrhoCDW})
of field theory apply. Here, the Bethe Ansatz 
solution~\cite{Cloizeaux,YangYang66} gives ($n_{\rm c}=1/2$,
$2 g(V)=h(\infty,V)$, $\cosh(\gamma)= V/V_{\rm c}$) 
\begin{eqnarray}
g(V) &=& 
\frac{1+2\sum_{n=1}^{\infty}(-1)^n/\cosh(n\gamma)}{1+2\sum_{n=1}^{\infty}
[1-\tanh(n\gamma)]}
\label{TaylorKrhoSF} \;, \\
g(V\to V_{\rm c}) &=&  \frac{2\pi}{\ln 2}
\exp\left(-\frac{\pi^2}{2\sqrt{2(V/V_{\rm c}-1)}}\right) \; .
\end{eqnarray}

As an application, we study the extended Hubbard model at quarter filling.
In Fig.~\ref{fig3}b, we show the phase diagram 
together with the contour lines for the TL~parameter $K_{\rho}$. 
Three different phases are found,
namely, a `superconducting phase' ($K_{\rho}>1$), a metallic
phase ($1/4 \leq K_{\rho} \leq 1$), and a 4$k_F$-CDW
insulator beyond the critical line.
The results are in good agreement with previous
works~\cite{Mil93,San04}. 

On the CDW transition line we find 
$K_{\rho}=1/4$, and $K_{\rho}^{\rm CDW}=1/8$ for
the infinitesimally doped CDW insulator,
in agreement with field theory~\cite{GiaSchulz}.
The parameter region where finite-size effects
are prominent due to an exponentially
small gap is shaded in Fig.~\ref{fig3}b.
Outside this region we can determine the TL~parameter
reliably. 

\begin{figure}[thb]
\vspace{18pt}
\begin{center}
\resizebox{9cm}{!}{\includegraphics{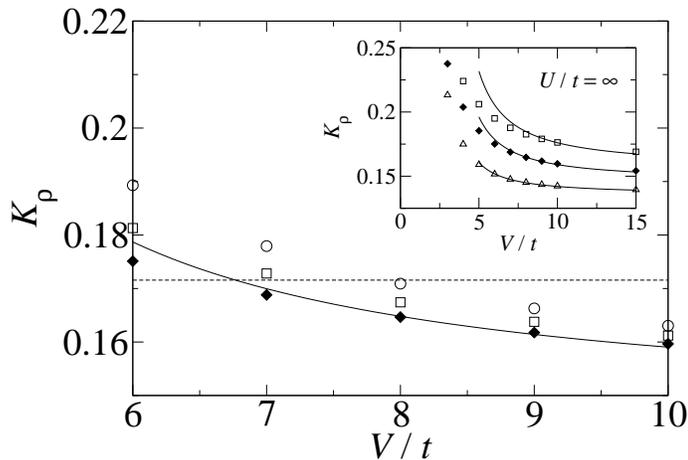}}
\end{center}
\caption{TL~parameter $K_{\rho}$ 
as a function of $V/t$ for $U/t=6,10,\infty$ at filling $n=11/24$
(from top to bottom).
The full line is the result~(\protect\ref{TaylorKrhoSF}),
the dashed horizontal line marks $K_{\rho}=3-2\sqrt{2}\approx 0.17$. 
Inset: $U/t=\infty$ for fillings 
$n=21/48,22/48,23/48$ (from top to bottom).\label{fig5}}
\end{figure}

The results for $\delta=1/24\approx 4\%$
are shown in Fig.~\ref{fig5} as a function of $V/t$ for $U/t=6,10,\infty$.
Deep in the CDW phase ($U/t\gtrsim 5$, $V/t\gtrsim 6$) neither $U$ nor~$V$
have a large influence on $K_{\rho}$.
Fig.~\ref{fig5} 
also shows, however, that $K_{\rho}<0.17$ and thus $\alpha>1$ requires 
very large interaction strengths, or very small doping, $\delta\lesssim 2\%$.

In conclusion, we developed an accurate numerical DMRG method 
to obtain the TL~parameter $K_{\rho}$ for Hubbard-type models. 
We demonstrated its accuracy for the 
Hubbard model and the XXZ Heisenberg chain (spinless fermions).
We presented an
accurate phase diagram of the $t$-$U$-$V$ model at quarter filling
and verified the field-theoretical predictions for~$K_{\rho}$.
We also showed that a critical exponent $\alpha>1$ is\ 
only possible
for a lightly doped CDW insulator with a sizable gap.

\acknowledgments
We thank E.~Jeckelmann,
R.M.~Noack, and F.H.L.~Essler for useful discussions.
S.E.~is supported by the Honjo International Scholarship Foundation.

\end{document}